\documentclass[Times,4pt,aps,pra,twocolumn,showpacs,amsmath,amssymb,floatfix,footinbib,superscriptaddress]{revtex4-1}
\usepackage{amsmath,natbib,graphicx,subfigure,amssymb,graphics,amsmath,mathrsfs,CJK,color}
\usepackage{multirow,fancyhdr,color,bm,tabularx,psfrag,dcolumn}
\usepackage[colorlinks=true,linkcolor=blue,urlcolor=blue,citecolor=blue]{hyperref}
\usepackage{tikz,xcolor,hyperref}% Make Orcid icon
\definecolor{lime}{HTML}{A6CE39}
\DeclareRobustCommand{\orcidicon}{%
    \begin{tikzpicture}
    \draw[lime, fill=lime] (0,0)
    circle [radius=0.16]
    node[white] {{\fontfamily{qag}\selectfont \tiny ID}};\draw[white, fill=white] (-0.0625,0.095)
    circle [radius=0.007];
    \end{tikzpicture}
    \hspace{-2mm}}
\foreach \x in {A, ..., Z}
{\expandafter\xdef\csname orcid\x\endcsname{\noexpand\href{https://orcid.org/\csname orcidauthor\x\endcsname}{\noexpand\orcidicon}}}

\begin{document}

%%\preprint{\emph{Submit to sss }} }
%%****************************************************************
\title{Charge-transport enhanced by the quantum entanglement in the Photosystem II reaction center }%\LARGE\boldmath\bf
%\thanks{}
\author{Ling-Fang Li }
\affiliation{Department of Physics, Faculty of Science, Kunming University of Science and Technology, Kunming, 650500, PR China}

\author{Shun-Cai Zhao\orcidA{}}
\email[Corresponding author: ]{zhaosc@kmust.edu.cn}
\affiliation{Department of Physics, Faculty of Science, Kunming University of Science and Technology, Kunming, 650500, PR China}

\author{Lu-Xin Xu }
\affiliation{Department of Physics, Faculty of Science, Kunming University of Science and Technology, Kunming, 650500, PR China}
%%****************************************************************
%\author{ }
%\affiliation{Department of Physics, Faculty of Science, Kunming University of Science and Technology, Kunming, 650500, PR China}
%\date{\today}

%%****************************************************************
%\linenumbers
\begin{abstract}
Revealing the role of quantum entanglement in charge-transport in the Photosystem II reaction center (PSII RC) is of great significance. In this work, we theoretically demonstrate that the robust quantum entanglement provides regulatory benefits to the charge-transport via a quantum heat engine (QHE) model with two absorbed photon channels. The calculation results manifest that the dynamic charge-transport and the steady-state photosynthetic properties of the PSII RC were enhanced by the intensity of quantum entanglement. Insight into the role of quantum entanglement in photosynthesis could motivate new experimental strategies for biomimetic photosynthetic devices in the future.
\begin{description}
\item[PACS numbers]42.50.Gy
\item[Keywords]{ charge-transport; quantum entanglement; Photosystem II reaction center }
\end{description}
\end{abstract}

\maketitle
\section{Introduction}

Recently, the mechanism of conversion efficiency close to one unit has attracted long-standing research interests\cite{Fleming2008Grand,Benniston2008Artificial,Ghosh2011Quantum} in photosynthesis, and more interests \cite{2013Quantum,Dorfman2013Photosynthetic,2015Natural} were attracted by the charge-transport in photosynthetic pigment-protein complexes for decades. To unravel the role of physical realm in Photosystem II reaction center (PSII RC), ultrafast optics and nonlinear spectroscopy technique\cite{Van2000Photosynthetic,2007Coherence,2002Prokhorenko} were utilizzed to the process of charge transport in light-harvesting complexes. In particular, evidence of quantum coherence\cite{2007Evidence} has been presented, with the idea that nontrivial quantum effects may be at the root of its remarkable efficiency. Savikhin et al. observed the quantum beating in the pigment-protein complexes via the fluorescence anisotropy technique\cite{Savikhin1997Oscillating}, as well as the long-lasting quantum beats provided direct evidence for survival of long-lived electronic coherence for hundreds of femtoseconds via the two-dimensional (2D) electronic spectra \cite{Engel2007Evidence,Duan8493} in the photosynthetic system, and the multi-dimensional spectroscopic measurements further demonstrated quantum coherences with timescales comparable to those of energy transfer processes\cite{0Quantum} in photosynthetic light-harvesting complexes.

Following these, some more theoretical and experimental works have attempted to reveal the precise role of quantum coherence in the light-harvesting complexes\cite{Van2000Photosynthetic,2007Coherence,2007Evidence,Miller2009Femtosecond,Collini2009Electronic,Panitchayangkoon2010Long} and have, perhaps surprisingly, found that environmental decoherence and noise plays a crucial role\cite{Dorfman2013Photosynthetic,2013Efficient,lizhao2021} in photosynthetic pigment-protein complexes. Dorfman et al.\cite{Dorfman2013Photosynthetic,lizhao2021} have verified that the noise-induced coherence in photosynthetic reaction centers, which is beneficial for increasing the photocurrent by at least $27\%$ owing to the quantum interference\cite{Marian2011Quantum}, and some other works demonstrated that the output power can be also improved by the quantum coherence in the photosynthetic system\cite{Svidzinsky2011Enhancing}.

More recently, the role of quantum coherence quantified by quantum entanglement, was demonstrated to exist in the process of excitation-energy transfer \cite{2007Quantum,Sarovar2010Quantum,PhysRevA.82.052109,PhysRevA.82.052109,Tan2012Quantum}, and the quantum entanglement was considered as a natural feature of coherent evolution\cite{2007Entanglement,2009Enhanced} due to the observed results in complex non-equilibrium chemical and biological processes. Therefore, some theoretical methods were introduced to quantify entanglement in the light-harvesting complex\cite{Sarovar2011Environmental}. Fassioli et al. showed the quantitative relationship between the entanglement distributions of the components of the Fenna-Matthew-Olson (FMO) complex\cite{Fassioli2010Distribution}. However, the sensitivity of the entanglement dynamics have not been directly evalued in the PSII RC, such as the dynamic energy flow and the steady-state photosynthetic properties via the QHE model\cite{Dorfman2013Photosynthetic, 2015Natural,2016Vibration,M2016A}.Then, theses factors breed the present work with double photon-absorbing channels. This work will analysis the dynamic charge-transport and steady-state photosynthetic properties dependent on the intensity of entanglement, and will try to
motivate new experimental strategies for biomimetic photosynthetic devices in the future.

\section{Physical Model and Solutions}

\subsection{Theoretical Model}

\begin{figure}[htp]
\subfigure[]{\includegraphics[scale=.5]{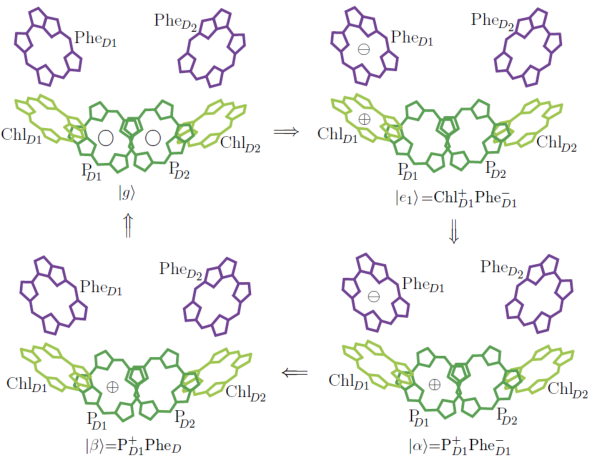}}
\hspace{0.30in}
\subfigure[]{\includegraphics[scale=.5]{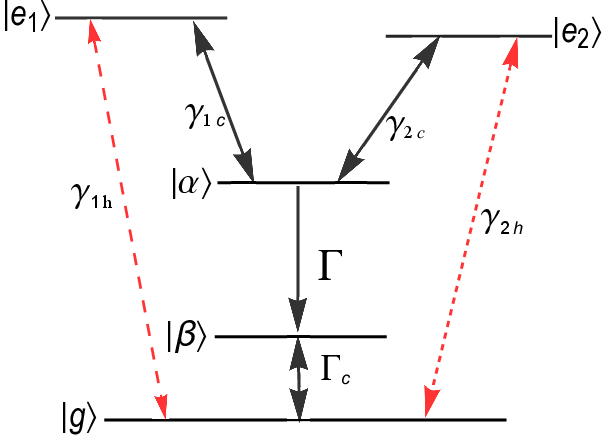}}\hspace{0in}%
\caption{( Color online ) (a) Four typical sates in the charge-transfer process (On the channel 1, $|g\rangle$$\leftrightarrow$$|e_{1}\rangle$$\leftrightarrow$$|\alpha\rangle$$\rightarrow$$|\beta\rangle$$\leftrightarrow$$|g\rangle$) of the six core-pigments in the PSII RC. \(|g\rangle\), six pigments in neutral ground state. \(|e_{1}\rangle\)=Chl\(^{+} _{D1}\)Phe\(_{D1}^{-}\) after Chl\(_{D1}\) primary charge-transfer state when the electron donor rapidly loses an electron to the nearby electron acceptor molecule Phe\(_{D1}\). \(|\alpha\rangle\)=P\(_{D1}^{+}\)Phe\(_{D1}^{-}\) charge-transfer state, after the positive and negative charges are spatially separated. \(|\beta\rangle\)=P\( _{D1}^{+}\)Phe\( _{D1}\), positively charged state, after an electron has been released from the system to perform work. (b) Corresponding five-level diagram of the photosynthetic QHE model with two photon-absorbing channels $|g\rangle$ to $|e_{1}\rangle$ and $|e_{2}\rangle$. $\gamma_{1h}$ and $\gamma_{2h}$ are the transition rates corresponding, $|\alpha\rangle$ is the charge separation state characterized by the transition rate $\gamma_{1h}(\gamma_{2h})$ from state $|e_{1}\rangle(|e_{2}\rangle)$ to state $|\alpha\rangle$, and the relaxation rate $\Gamma$ to state $|\beta\rangle$, $\Gamma_{C}$ is the relaxation rate to the ground state $|g\rangle$ of the system.}
\label{Fig.1}
\end{figure}

In order to analysis the effect of the quantum entanglement on charge-transport performance, we introduce a QHE model to describe the charge transfer process\cite{2007Photosystem,2011Light,2010Two} in
Fig.\ref{Fig.1}.  The typical PSII-RC found in purple bacteria and in oxygen-evolving organisms contains six pigment molecules aligned in two branches\cite{2015Multiple}, which contains four chlorophylls (special
pair P\(_{D1}\) and P\(_{D2}\) and accessory Chl\(_{D1}\) and Chl\(_{D2}\)) and two pheophytins (Phe\(_{D1}\) and Phe\(_{D1}\)) arranged in two branches\cite{M2017Effects}.

As shown in Fig.\ref{Fig.1} (a), \(|g\rangle\) denotes all six pigments in neutral ground state after an electron has been replenished at the special pair. \(|e_{1}\rangle\)=Chl\(^{+} _{D1}\)Phe\(_{D1}^{-}\) describes primary charge-transfer state after Chl\(_{D1}\)  when the electron donor rapidly loses an electron to the nearby electron acceptor molecule Phe\(_{D1}\). $|\alpha\rangle$  represents charge-transfer state P\(_{D1}^{+}\)Phe\(_{D1}^{-}\) with the positive and negative charges spatially separated. And \(|\beta\rangle\) expresses positively charged state, P\( _{D1}^{+}\)Phe\( _{D1}\) after an electron has been released out of the system to perform work. Fig.\ref{Fig.1} (b) shows a five-level photosynthetic QHE model corresponding to Fig.\ref{Fig.1} (a). The charge transfer initiates the photon-absorbing from $|g\rangle$ to $|e_{1}\rangle$ and $|e_{2}\rangle$ with transition rates $\gamma_{1h}$ and $\gamma_{2h}$, respectively. And the charge-separated processes $|e_{1}\rangle$$\leftrightarrow$$|\alpha\rangle$$\leftrightarrow$$|\beta\rangle$, $|e_{2}\rangle$$\leftrightarrow$$|\alpha\rangle$$\leftrightarrow$$|\beta\rangle$ take place at the rate $\gamma_{1c}$ and $\gamma_{2c}$, respectively. The quantum entanglement may arise in these charge-separated processes. Physically, the quantum entanglement is a natural feature of coherent evolution. Therefore, the entanglement dynamics in this PSII RC can be considered a quantum interference effect\cite{Tan2012Quantum} between above two charge-separated processes can be described by two charge-transfer states, \(|e_{1}\rangle\)=Chl\(^{+} _{D1}\)Phe\(_{D1}^{-}\) and \(|e_{2}\rangle\)=Chl\(^{+} _{D2}\)Phe\(_{D2}^{-}\), which starts with the initial state $|\Psi(0)\rangle$=$\frac{1}{\sqrt{2}}(|e_{1}\rangle-|e_{2}\rangle)$. For a bipartite entanglement between the two charge-transfer states, the standard measure for quantum entanglement, the concurrence\cite{Sarovar2010Quantum,1997Entanglement} is employed as follows,

\begin{equation}
\mathcal{C}_{12}=2|\rho_{e_{1}e_{2}}|,
\end{equation}

\noindent During above two processes, the total photosynthetic power current $j=e\Gamma\rho_{\alpha\alpha}$ formed by the performing work at a rate $\Gamma$, drives the chain of chemical reactions from $|\alpha\rangle$ to $|\beta\rangle$ with $e$ being the elementary electron charge. Finally, the electron returns to the neutral ground state $|g\rangle$ at the rate of  $\Gamma_{C}$ and emits phonons to complete the cycle.

With these knowledge, the integral Hamiltonian for the PSII RC plus the ambient environment can be represented by

\begin{align}
\hat{H}=\hat{H}_{S}+\hat{H}_{B}+\hat{H}_{SB},
\end{align}

\noindent where the electronic Hamiltonian with the energy $E_{i}$ of the ith state is read as
\begin{eqnarray}
& \hat{H}_{S}=&E_{0}|g\rangle\langle g|+E_{e_{1}}|e_{1}\rangle\langle e_{1}|+E_{e_{2}}|e_{2}\rangle\langle e_{2}|\nonumber\\
&&+E_{\alpha}|\alpha\rangle\langle \alpha|+E_{\beta}|\beta\rangle\langle \beta|,
\end{eqnarray}

\noindent Among Eq.(2), $\hat{H}_{B}$ describes the Hamiltonian of the ambient thermal bath. Considering the weak coupling between the PSII RC and the ambient environment, the harmonic oscillator thermal bath model is generally introduced to describe the ambient environment as follows,

\begin{align}
\hat{H}_{B}=\sum_{k}\hbar\omega_{k}\hat{a}^{\dag}_{k}\hat{a}_{k},
\end{align}

\noindent where $\hat{a}^{\dag}_{k}(\hat{a}_{k})$ are the creation (annihilation) operator of the kth harmonic oscillator mode with its frequency \(\omega_{k}\). In Eq.(2), $\hat{H}_{SB}$ describes the interaction Hamiltonian between the corresponding ambient reservoir modes and the PSII RC. Under the rotating-wave approximations,  $\hat{H}_{SB}$ is expressed as follows,

\begin{align}
&\hat{H}_{SB}=\hat{V}_{h} + \hat{V}_{c}\nonumber, \\
&\hat{V}_{h}=\sum_{i=1,2}\sum_{k}\hbar(\varepsilon_{ik}\hat{\sigma}_{gi}\otimes\hat a_{k}^{\dag}+\varepsilon_{ik}^{\ast}\hat{\sigma}_{gi}^{\dag}\otimes\hat a_{k})\nonumber, \\
&\hat{V}_{c}=\sum_{j=1,2}\sum_{k}\hbar(\varepsilon_{jk}\hat{\sigma}_{\alpha j}\otimes\hat b_{k}^{\dag}+\varepsilon_{jk}^{\ast}\hat{\sigma}_{\alpha j}^{\dag}\otimes\hat b_{k}). \\\nonumber
\end{align}

\noindent where $\varepsilon_{ik}(\varepsilon_{jk})$ is the corresponding coupling-strength between the i(j)th\(_{(i,j=1,2)}\) charge transfer process and the kth mode of ambient environment reservoir, and the transit operators are defined as $\hat{\sigma}_{gi}=|g\rangle\langle e_{i}|$, $\hat{\sigma}_{\alpha j}$=$|\alpha\rangle\langle e_{j}|_{(i,j=1,2)}$ with $\hat{a}^{\dag}_{k}$, $\hat{b}^{\dag}_{k} (\hat{a}_{k}$, $\hat{b}_{k})$ being the creation(annihilation) operators of the kth reservoir mode, respectively.

\subsection{Master equations for the PSII RC}

Under the Born-Markov and Weisskopf-Wigner approximations in the Schr$\ddot{o}$dinger picture, the PSII RC can be described by the master equation deduced by the conventional second-order perturbative treatment for $\hat{V}_{h}$ and $\hat{V}_{c}$. Taking the Lindblad-type superoperators, the master equation is read as,

\begin{equation}
\frac{d\hat{\rho}}{dt}=-i[\hat{H}_{S},\hat{\rho}]+\sum_{i,j=1,2}\mathscr{L}_{ijh}\hat{\rho}+\mathscr{L}_{ijc}\hat{\rho}+\mathscr{L}_{\Gamma_{C}}\hat{\rho}+\mathscr{L}_{\Gamma}\hat{\rho},
\end{equation}

\noindent In Eq.(6), the first term on the right side describes the coherent evolution of the PSII RC, and the other four Lindblad-type superoperator terms are deduced with the following expressions,

\begin{widetext}
\begin{eqnarray}
&\mathscr{L}_{ijh}\hat{\rho}=&\sum_{i,j=1,2}\frac{\gamma_{ijh}}{2}[(n_{ih}+1)(\hat{\sigma}_{gi}\hat{\rho}\hat{\sigma}_{gj}^{\dag}+\hat{\sigma}_{gj}\hat{\rho}\hat{\sigma}_{gi}^{\dag}
-\hat{\sigma}_{gj}^{\dag}\hat{\sigma}_{gi}\hat{\rho}-\hat{\rho}\hat{\sigma}_{gj}^{\dag}\hat{\sigma}_{gi})\nonumber\\
&&+n_{ih}(\hat{\sigma}_{gi}^{\dag}\hat{\rho}\hat{\sigma}_{gj}+\hat{\sigma}_{gj}^{\dag}\hat{\rho}\hat{\sigma}_{gi}
-\hat{\sigma}_{gj}\hat{\sigma}_{gi}^{\dag}\hat{\rho}-\hat{\rho}\hat{\sigma}_{gj}\hat{\sigma}_{gi}^{\dag})],\\
&\mathscr{L}_{ijc}\hat{\rho}=&\sum_{i,j=1,2}\frac{\gamma_{ijc}}{2}[(n_{ic}+1)(\hat{\sigma}_{\alpha i}\hat{\rho}\hat{\sigma}_{\alpha j}^{\dag}
+\hat{\sigma}_{\alpha j}\hat{\rho}\hat{\sigma}_{\alpha i}^{\dag}-\hat{\sigma}_{\alpha j}^{\dag}\hat{\sigma}_{\alpha i}\hat{\rho}-\hat{\rho}\hat{\sigma}_{\alpha j}^{\dag}\hat{\sigma}_{\alpha i})\nonumber\\
&&+n_{ic}(\hat{\sigma}_{\alpha i}^{\dag}\hat{\rho}\hat{\sigma}_{\alpha j}+\hat{\sigma}_{\alpha j}^{\dag}\hat{\rho}\hat{\sigma}_{\alpha i}
-\hat{\sigma}_{\alpha j}\hat{\sigma}_{\alpha i}^{\dag}\hat{\rho}-\hat{\rho}\hat{\sigma}_{\alpha j}\hat{\sigma}_{\alpha i}^{\dag})],\\
&\mathscr{L}_{\Gamma_{C}}\hat{\rho}=&\frac{\Gamma_{C}}{2}[(N_{C}+1)(2\hat{\sigma}_{g\beta}\hat{\rho}\hat{\sigma}_{g\beta}^{\dag}-\hat{\sigma}_{g\beta}^{\dag}\hat{\sigma}_{g\beta}
\hat{\rho}-\hat{\rho}\hat{\sigma}_{g\beta}^{\dag}\hat{\sigma}_{g\beta})\nonumber\\
&&+N_{C}(2\hat{\sigma}_{\beta g}^{\dag}\hat{\rho}\hat{\sigma}_{\beta g}-\hat{\sigma}_{\beta g}\hat{\sigma}_{\beta g}^{\dag}\hat{\rho}-\hat{\rho}\hat{\sigma}_{\beta g}\hat{\sigma}_{\beta g}^{\dag})],\\
&\mathscr{L}_{\Gamma}\hat{\rho}=&\frac{\Gamma}{2}(2\hat{\sigma}_{\beta\alpha}\hat{\rho}\hat{\sigma}_{\beta\alpha}^{\dag}-\hat{\sigma}_{\beta\alpha}^{\dag}\hat{\sigma}_{\beta\alpha}\hat{\rho}
-\hat{\rho}\hat{\sigma}_{\beta\alpha}^{\dag}\hat{\sigma}_{\beta\alpha})
\end{eqnarray}
\end{widetext}

\noindent  $\mathscr{L}_{ijh}\hat{\rho}$ in the expression (7) denotes the dissipative effect of the ambient environment photon reservoirs.  And $\gamma_{iih}$=$\gamma_{ih}$, $\gamma_{jjh}$=$\gamma_{jh}$ are the spontaneous decay rates from the state \(|e_{i}\rangle(i$=$1,2)\) to the neutral ground state \(|g\rangle\) respectively. $\gamma_{ijh}$=$\gamma_{jih}$, the cross-coupling describes the effect of Fano interference. It is assumed that \(\gamma_{ijh}$=$\sqrt{\gamma_{ih}\gamma_{jh}}(i,j$=$1,2)\)  represents the maximal interference and \(\gamma_{ijh}$=$0\) for the minimal interference. And $n_{ih}(i$=$1,2)$ denotes the average electron occupations on the corresponding charge separation state.  The expression $\mathscr{L}_{ijc}\hat{\rho}$  describes the effect of the low temperature reservoirs with the average phonon numbers \(n_{ic}$=$[exp(\frac{(E_{e_{i}}-E_{\alpha})}{k_{B}T_{a}})-1]^{-1}(i$=$1,2)\) and \(T_{a}\) being the ambient temperature. $\gamma_{iic}$=$\gamma_{ic}$, $\gamma_{jjc}$=$\gamma_{jc}$ are the corresponding spontaneous decay rates from the state \(|e_{i}\rangle(i$=$1,2)\) to state \(|\alpha\rangle\). $\gamma_{ijc}$ is the cross-coupling that represents the effect of Fano interference, which is defined by $\gamma_{ijc}$=$\gamma_{jic}$$=$\(\gamma_{ijc}$=$\sqrt{\gamma_{ic}\gamma_{jc}}(i,j$=$1,2)\) with fully interference while \(\gamma_{ijc}=0\) for no interference. Similarly, $\mathscr{L}_{\Gamma_{C}}\hat{\rho}$ in expression (9) describes another interaction between the PSII RC and ambient environment with $\hat{\sigma}_{g\beta}=|\beta\rangle\langle g|$, where \(N_{C}$=$[exp(\frac{(E_{\beta}-E_{g})}{k_{B}T_{a}})-1]^{-1}\) denotes the corresponding average phonon occupation numbers. The last term $\mathscr{L}_{\Gamma}\hat{\rho}$ describes a process that the PSII RC in state \(|\alpha\rangle\) decays to state $|\beta\rangle$. It leads to the photosynthetic power current proportional to the relaxation rate $\Gamma$ as defined before, and the operator $\hat{\sigma}_{\beta\alpha}$ is defined as $\hat{\sigma}_{\beta\alpha}$=$|\beta\rangle\langle\alpha|$. Thereupon, the elements of the density matrix $\rho$ can be written as follows,

\begin{widetext}
 \begin{align}
\dot{\rho}_{e_{1}e_{1}}={}&-\gamma_{1h}[(n_{1h}+1)\rho_{e_{1}e_{1}}-n_{1h}\rho_{gg}]-\gamma_{1c}[(n_{c}+1)\rho_{e_{1}e_{1}}-n_{c}\rho_{\alpha\alpha}]
                        \nonumber\\{}&-\gamma_{12h}[(n_{1h}+1) Re[\rho_{e_{1}e_{2}}]-\gamma_{12c}[(n_{c}+1) Re[\rho_{e_{1}e_{2}}],\nonumber\\
\dot{\rho}_{e_{2}e_{2}}={}&-\gamma_{2h}[(n_{2h}+1)\rho_{e_{2}e_{2}}-n_{1h}\rho_{gg}]-\gamma_{2c}[(n_{c}+1)\rho_{e_{2}e_{2}}-n_{c}\rho_{\alpha\alpha}]
                        \nonumber\\{}&-\gamma_{12h}[(n_{2h}+1)Re[\rho_{e_{1}e_{2}}]-\gamma_{12c}[(n_{c}+1) Re[\rho_{e_{1}e_{2}}],\nonumber\\
\dot{\rho}_{\alpha\alpha}={}&-\gamma_{1c}[(n_{c}+1)\rho_{e_{1}e_{1}}-n_{c}\rho_{\alpha\alpha}]+\gamma_{2c}[(n_{c}+1)\rho_{e_{2}e_{2}}-n_{c}\rho_{\alpha\alpha}]
                        \nonumber\\{}&+2\gamma_{12c}[(n_{c}+1)Re[\rho_{e_{1}e_{2}}]-\Gamma\rho_{\alpha\alpha},\\
\dot{\rho}_{\beta\beta}={}&\Gamma\rho_{\alpha\alpha}-\Gamma_{C}[(\emph{N}_{C}+1)\rho_{\beta\beta}+\emph{N}_{C}\rho_{gg}],\nonumber\\
\dot{\rho}_{e_{1}e_{2}}={}&-\frac{\gamma_{1h}}{2}(n_{1h}+1)\rho_{e_{1}e_{2}}-\frac{\gamma_{2h}}{2}(n_{2h}+1)\rho_{e_{1}e_{2}}-\frac{\gamma_{12h}}{2}[(n_{1h}+1)\rho_{e_{1}e_{1}}+(n_{2h}+1)\rho_{e_{2}e_{2}}
              \nonumber\\{}&-n_{1h}\rho_{gg}-n_{2h}\rho_{gg}]-\frac{\gamma_{1c}+\gamma_{2c}}{2}(n_{c}+1)\rho_{e_{1}e_{2}}
              \nonumber\\{}&-\frac{\gamma_{12c}}{2}[(n_{c}+1)\rho_{e_{1}e_{1}}+(n_{c}+1)\rho_{e_{2}e_{2}}-2n_{c}\rho_{\alpha\alpha}].\nonumber
\end{align}
\end{widetext}

\noindent  where $\rho_{ii}$ describe the diagonal element and $\rho_{ij}$ is the non-diagonal element of the corresponding states.

\section{Results and discussion}

Instead of the investigation to the entanglement time evolution for various subsystems of the qubit network modeling the FMO complex\cite{Caruso2010Entanglement}, this work devotes into the influence of the intensity of  entanglement on the photosynthetic performance, which will be quantitatively evaluated by the input power, population dynamics of the charge separate sate, j-V characteristic and output power. We recognize that the energy flux entering the PSII RC can be randomly modulated by the quantum entanglement through the two photon-absorbing channels. Then, a simple probability and power condition is introduced to measure the input power $U_{m}$\cite{2015Natural},

\begin{equation}
\rho_{e_{1}e_{1}} U_{1}+\rho_{{e_{2}e_{2}}}U_{2}=U_{m}
\end{equation}

\noindent  where $U_{1}=E_{1}n_{1h}$ and $U_{2}=E_{2}n_{2h}$ with $E_{1}$ and $E_{2}$ are the excitation energies of the two states \(|e_{1}\rangle\), \(|e_{2}\rangle\). And the voltage corresponding to the current is defined as the chemical potential difference between the two states $|\alpha\rangle$ and $|\beta\rangle$, which can be expressed as $eV=E_{\alpha}-E_{\beta}+k_{B}T_{a}\ln(\frac{\rho_{\alpha\alpha}}{\rho_{\beta\beta}})$ with the steady-state solutions to Eq.(11) by the Boltzmann distribution. Then, the output power can be represented as $P$=$j\cdot V$ with the selected parameters from Table \ref{Table 1} for the proposed theoretical model.

\begin{table}
\begin{center}
\caption{Model parameters used in the numerical calculations.}
\label{Table 1}
\vskip 0.2cm\setlength{\tabcolsep}{0.5cm}
\begin{tabular}{ccc}%{|c|c|c|c|c|c|c|c|c|c|}
\hline
\hline
                                                                        & Values                 & Units  \\
\hline
\(E_{1}-E_{\alpha}\)                                                    & 0.2                    & eV  \\
\(E_{\beta}-E_{g}\)                                                     & 0.2                    & eV  \\
\(\gamma_{1h}\)                                                         & 0.62                   & eV  \\
\(\gamma_{2h}\)                                                         & 0.56                   & eV  \\
\(\gamma_{1c}\)                                                         & 0.22                   & eV  \\
\(\gamma_{2c}\)                                                         & 0.82                   & eV  \\
\(\Gamma\)                                                              & 0.56                   & eV  \\
\(\Gamma_{C}\)                                                          & 0.68                   & eV  \\
\(n_{1h}\)                                                              & 60                     &     \\
\(n_{2h}\)                                                              & 58                     &     \\
\(T_{a}\)                                                               & 300                    & K   \\
\hline
\hline
\end{tabular}
\end{center}
\end{table}

\begin{figure}[htp]
\center
\includegraphics[scale=.5]{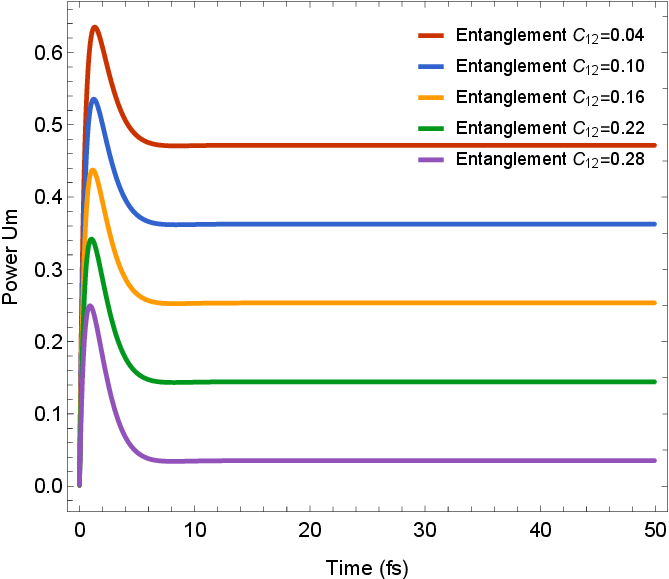 }
\caption{(Color online) Dynamics evolution of the input power $U_{m}$ influenced by different entanglement intensities with other parameters taken from Table \ref{Table 1}. }
\label{Fig.2}
\end{figure}

In the following, the charge-transport performance dependent on the intensity of entanglement catches our interesting. In this context, the input power, population dynamics of the charge-transfer state sate, j-V characteristic and output power are proposed to evaluated the photosynthetic performance, which will be modulated by the intensity of entanglement. Firstly, the dynamics evolution of the input power $U_{m}$ was numerically plotted in Fig.\ref{Fig.2} with different entanglement intensities $\mathcal{C}_{12}$ =0.04, 0.10, 0.16, 0.22 and 0.28.

As shown in Fig.\ref{Fig.2}, the curves ascend to the peak values then descend quickly, and sequently shows the horizontal evolution characteristics in the time range of [0, 50fs], and the values of input power $U_{m}$ decrease with the increment of $\mathcal{C}_{12}$ in the time range of [0, 50fs]. The physical significance in Fig.\ref{Fig.2} can be deduced from Eq.(12), which shows the value of Eq.(12) is proportion to the populations on the two charge-transfer states $|e_{1}\rangle$ and $|e_{2}\rangle$. The smaller $U_{m}$ indicates less charges populated on the two charge-transfer states $|e_{1}\rangle$ and $|e_{2}\rangle$, which may mean two opposite possibilities of the charge populations. One is less charges transferred to the two charge-transfer states $|e_{1}\rangle$ and $|e_{2}\rangle$ due to the negative influence caused by the quantum entanglement, which may causes less efficient charge transport. If we turn our attention to the charge-transfer state \(|\alpha\rangle\)=P\(_{D1}^{+}\)Phe\(_{D1}^{-}\) in Fig.\ref{Fig.1}(b), which is in the next stage of the two states $|e_{1}\rangle$, $|e_{2}\rangle$ in the total charge-transport process. Thence, the other possibility is the higher charge-transport efficiency owing to more charges transferred to next charge-transfer state $|\alpha\rangle$. Therefore, it's hard to draw a conclusion about the explicit role of quantum entanglement in the charge-transport efficiency from Fig.\ref{Fig.2}.

\begin{figure}[htp]
\center
\includegraphics[scale=.5]{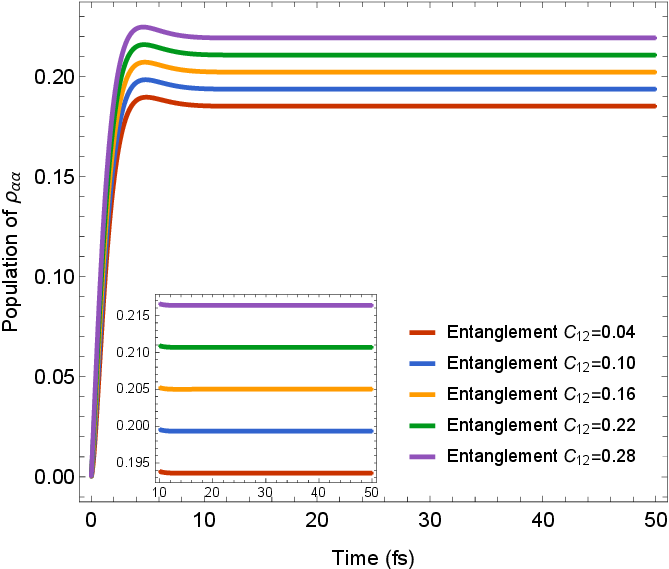 }
\caption{(Color online) Population dynamics of the state \(|\alpha\rangle\)=P\(_{D1}^{+}\)Phe\(_{D1}^{-}\) with different quantum entanglement intensities  at the room temperature. Other parameters are the same to those in Fig.\ref{Fig.2}.}
\label{Fig.3}
\end{figure}

With the mentioned topic in mind, we investigate the population dynamics of the charge-transfer state \(|\alpha\rangle\)=P\(_{D1}^{+}\)Phe\(_{D1}^{-}\). The populations on state \(|\alpha\rangle\)
is illustrated in Fig.\ref{Fig.3} with the dynamic solutions to Eq.(11) and the same parameters to Fig.\ref{Fig.2}. The curves in the range of [0, 50fs] and its inset in the range of [10, 50fs] in Fig.\ref{Fig.3} clearly display that the populations on state \(|\alpha\rangle\) increase with the bipartite quantum entanglement in the time evolution. It means more charges transported to the state \(|\alpha\rangle\) owing to the intensities of entanglement.
This result answers the mentioned question raised by Fig.\ref{Fig.2}, i. e., more charges are transported to the state \(|\alpha\rangle\), and that brings about less populations on the charge-transfer states $|e_{1}\rangle$ and $|e_{2}\rangle$. It comes to that an efficietn the charge-transfer is achieved by the increasing entanglement intensities $\mathcal{C}_{12}$ owing to the positive role of $\mathcal{C}_{12}$ in the PSII RC.

\begin{figure}[htp]
\center
\includegraphics[scale=.5]{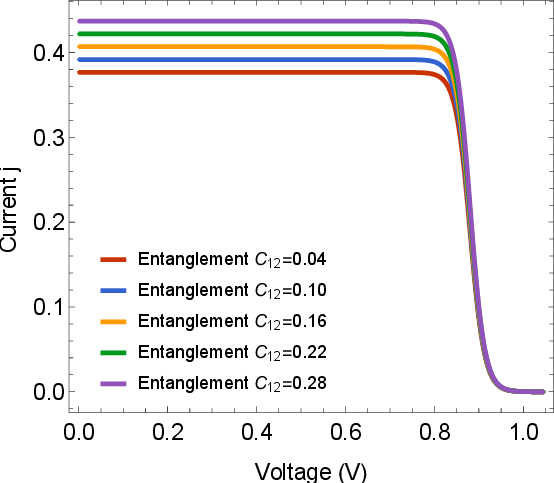 }
\caption{(Color online) Current $j$ as the function of effective voltage $V$ at room temperature with different quantum entanglement intensities  at the room temperature. Other parameters are taken from Table \ref{Table 1}.}
\label{Fig.4}
\end{figure}

\begin{figure}[htp]
\center
\includegraphics[scale=.5]{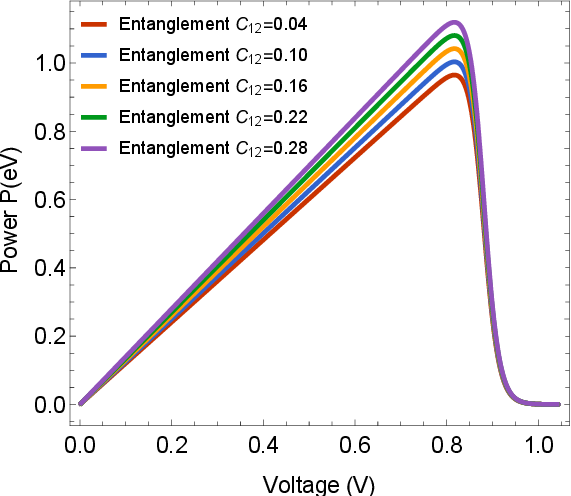 }
\caption{(Color online) The output power $P$ as a function of effective voltage $V$ at room temperature with different quantum entanglement intensities at the room temperature.  Other parameters are the same to Fig.\ref{Fig.4}.}
\label{Fig.5}
\end{figure}

As mentioned above, in PSII RC, the obtained dynamic photosynthetic properties are enouraging, while its steady-state photosynthetic characteristics is also worth of our attention. Therefore, the steady-state solutions to Eq.(11) were performed for the current j and the output power $P$ in Fig.\ref{Fig.4} and Fig.\ref{Fig.5}, respectively. As shown by the curves in Fig.\ref{Fig.4}, the short-circuit currents $j$ increase with entanglement intensities $\mathcal{C}_{12}$. It means that more photosynthetic energ will be exported by the PSII RC, and more efficient charge-transport efficiency was achieved when the PSII RC is accompanied by robust entanglement with two absorbing-photon channels.

Another key physical quantity, i.e., the output power $P$ is often utilized to valuate the photosynthetic property in the process of photosynthesis. Then, in Fig.\ref{Fig.5} we plot the output power $P$ versus the effective voltage $V$, and we still take the bipartite quantum entanglement $\mathcal{C}_{12}$ = 0.04, 0.10, 0.16, 0.22 and 0.28. The curves clearly show that the peak power $P$ is gradually enhanced by the increment of quantum entanglement intensity $\mathcal{C}_{12}$ when the PSII RC oprates in the steady-state process. Then, this result in Fig.\ref{Fig.5} once again demonstrates that the photosynthetic properties are enhanced by the bipartite entanglement $\mathcal{C}_{12}$ between two charge-transport states in the PSII RC.

Before concluding this paper, we would like to give some remarks on the above obtained results. Firstly, our current work focuses on the intensity of entanglement in the PSII RC, even though the influence of the entanglement was discussed both in the time evolution and in the steady-state, and some significant results were achieved in the PSII RC. Nevertheless, the effect of lifetime of entanglement on the charge-transport in the PSII RC is not mentioned in this work. Recent work shows that the two-dimensional (2D) electronic spectra technique is widely used to measure the lifetimes of quantum coherence\cite{Duan8493,0Quantum,doi:10.1063/1.5119248}. Thus, some more interesting physical regime may be revealed when the entanglement lifetime is considered in our further work.

Secondly, the above obtained results could actually be seen as providing some potential experimental tests for efficient photosynthetic performance in PSII RC. However, considering its sensitivity and fragility to the precise evolution of the environment, accurate measurement of entanglement maybe difficult for an experiment in PSII RC. In order to solve this problem, some theoretical schemes for valuating the intensity of entanglement should be available for comparison, and in this respect the nature of the environment is central to the problem of charge-transport in PSII RC. Further studies are, however, required before any result can be accepted as conclusive on the functional and possibly beneficial role of entanglement in charge-transport in PSII RC.

\section{Conclusions}

In conclusion, special attention is paid to the role of the intensity of entanglement in photosynthetic performance. We investigated the influence of quantum entanglement in the PSII RC, which is described by QHE model with two absorbing-photon channels. The results display that the charge-transport efficiency evaluated by the input power and dynamic population on the charge-transfer state can be enhanced by the quantum entanglement intensity in the time evolution. Moreover, we have found that the photosynthetic performance characterized by current and output power can also benefit from quantum entanglement intensity under the steady condition.

\section{acknowledgments}

We offer our thanks for the financial support from the National Natural Science Foundation of China ( Grant Nos. 62065009 and 61565008 ), and Yunnan Fundamental Research Projects, China ( Grant No. 2016FB009 ).

\section*{Conflict of Interest}

The authors declare that they have no conflict of interest. This article does not contain any studies with human participants or animals performed by any of the authors. Informed consent was obtained from all individual participants included in the study.

\bibliography{reference}
\bibliographystyle{unsrt}%abbrv, acm, alpha, apalike, ieeetr, plain, siam,unsrt.
\end{document}